\documentstyle[11pt,psfig,epsf]{article}

\def\be{\begin{eqnarray}}
\def\ee{\end{eqnarray}}
\def\bq{\begin{equation}}
\def\eq{\end{equation}}
\def\ben{\begin{enumerate}}\def\een{\end{enumerate}}
\def\d {\partial}
\def\L{{\cal L}}
\def\mn{m_N}
\def\mnstar{m_N^\star}\def\Mnstar{M_N^\star}\def\kF{k_F}
\def\mpi{m_\pi}
\def\F1{F_1}
\def\ft1{{\tilde{F}_1}}
\def\ft1omega{\tilde{F}_1^\omega}

\def\prl {Phys. Rev. Lett.}\def\pr{Phys. Rev.}
\def\np{Nucl. Phys.}\def\pl{Phys. Lett.}
\def\la{\langle}\def\ra{\rangle}
\def\O{{\cal O}}
\def\roughly#1{\mathrel{\raise.3ex\hbox{$#1$\kern-.75em%
\lower1ex\hbox{$\sim$}}}}

\def\gsim{\roughly>}
\def\f{{f_\pi^\star}}
\def\barnu{\bar{\nu}}
\def\mec{\epsilon_{MEC}}
\begin{document}
\begin{titlepage}
\hfill
\renewcommand{\thefootnote}{\fnsymbol{footnote}}
\vspace{.2cm}
\begin{center}
\ \\
{\large \bf CHIRAL SYMMETRY}

{\large \bf AND}

{\large \bf MESON-EXCHANGE CURRENTS}\footnote{Based on invited talks given at 1997 Joint APS/AAPT Washington meeting, 18-21 April 1997, at the Worskhop on ``Hadrons in Dense Matter,"
GSI, Darmstadt (Germany), 2-4 July 1997 and at the Workshop on ``Hadron 
Systems in High Density and/or High Temperature," 
Argonne National Lab. (USA), 4-8 August 1997.}
\ \\
\ \\
\vspace{1.0cm}
{Mannque Rho}
\vskip 0.5cm

{\it Service de Physique Th\'eorique, CEA Saclay,}

{\it F-91191 Gif-sur-Yvette, France}
\vskip 0.3cm

{(27 August 1997)}

\end{center}
\vskip 0.6cm
\centerline{\bf ABSTRACT}
\vskip 0.6cm
The meson-exchange current in nuclei, a long-standing problem in nuclear
physics, is described in modern theory of strong interactions,
namely, QCD expressed in terms of effective chiral Lagrangian field theory.
Some old results are given a modern interpretation and some new results are
predicted. Among the topics treated are an accurate calculation of the 
radiative np
capture process at thermal energy, the enhanced axial charge transition in
heavy nuclei, Brown-Rho (BR) scaling in dense medium induced by vacuum changes,
dropping meson masses and ``mended symmetry" in relativistic heavy-ion processes 
and the link
between the physics of dilute and dense hadronic systems through a
mapping of effective chiral Lagrangians to Landau Fermi-liquid theory.

%\vspace{2cm}

\end{titlepage}
\renewcommand{\thefootnote}{\arabic{footnote}}

In a recent {\it Nature} article, Weinberg, in summarizing the present 
understanding of what is meant by ``elementary particles" and field theory,
wrote \cite{weinnature}:
{\it We have come to understand that particles may be described at sufficiently
low energies by fields appearing in so-called effective quantum field theories,
whether or not these particles are truly elementary. For instance,
even though nucleon
and pion fields do not appear in the Standard Model, we can calculate the
rates for processes involving low-energy pions and nucleons by using an 
effective quantum field theory of pion and nucleon fields rather than of 
quark and gluon fields. .... When we use a field theory in this way, we 
are simply invoking the general principles of relativistic quantum theories,
together with any relevant symmetries; we are not making any assumption about
the fundamental structure of physics.}

In this article, I would like to illustrate how this statement can be applied
to nuclear physics. I will start with a well-known case of the lightest
nucleus, namely, two-nucleon system, go to heavy nuclei and then finally to
dense matter with a density higher than nuclear matter density.

Consider a process involving two-nucleon systems. The classic case studied
for almost half a century is the process
\be
n+p\rightarrow d +\gamma \label{np}
\ee
at thermal energy, with the relative momentum in the center of mass system
$p\simeq 3.4451\times 10^{-5}$ MeV. This process was explained within
10\% accuracy by Austern already in 1953 \cite{austern}
and the remaining 10\% discrepancy
was explained in terms of meson-exchange currents by Riska and Brown in 
1974 \cite{riskabrown}.
I will now describe how one can completely understand this process in
an effective chiral Lagrangian formalism \cite{pmr}.

For this process, we can start 
with a theory defined in the vacuum, that is to say, in  matter-free
space since the two-body system is a dilute one. From the
QCD point of view, the essential physics is dictated by the quark condensate
in the vacuum $\la \bar{q}q\ra$
since apart from the small up and down quark masses, the masses and 
couplings of the relevant degrees of freedom, i.e., light-quark hadrons, are dictated by the spontaneously broken
chiral symmetry. This is because the length scale is set by the quark condensate
in the vacuum. Thus the relevant theory is the one that one should be able
to write down in matter-free space. 

Now following the general strategy of
effective field theories, we first have to identify the relevant degrees 
of freedom that we would like to treat explicitly and put all irrelevant
degrees of freedom into the constants appearing in the Lagrangian.
Since the process involved (\ref{np}) is a very low-energy process,
we take explicitly as relevant degrees of freedom 
the proton and neutron fields denoted as a doublet $N$
together with the
Goldstone excitations of spontaneously broken chiral symmetry,
namely, the pions $\pi^i$. For the moment, other heavy degrees of freedom 
such as
the baryon resonance $\Delta$, the vector mesons $\rho$ and $\omega$
and the scalar $\sigma$ will be integrated out so that they will not figure
explicitly in the theory. They will of course figure somewhere in the theory
and I will show where later.

The next step is to write down the most general Lagrangian consisting of
the $N$ and $\pi^i$ fields consistent with the general properties Weinberg
is referring to above,  notably, spontaneously
broken chiral symmetry. The Lagrangian will contain, in addition to
bilinears in the $N$ field, terms involving $4N$, $6N$ etc. to all orders
in power of fermion fields suitably coupled to Goldstone pions
and since light-quark masses are not zero, though small, there should be
terms involving quark mass terms. In nuclear physics at low energies,
the nucleon can be considered as heavy. When the nucleon is treated
in that way, 
the leading Lagrangian, when expanded, can be written as
\be
\L&=& N^\dagger \left(i\d^0 +\frac{\nabla^2}{2M}\right)N -\frac{g}{f_\pi}
\nabla {\bf \pi}\cdot N^\dagger{\bf \tau} \sigma N +\frac 12 (\d^0{\bf \pi})^2 -
\frac 12(\nabla{\bf \pi})^2 -\frac 12 m_\pi^2 ({\bf \pi})^2 \nonumber\\
&& + \frac{C_1}{f_\pi^2} (N^\dagger N)^2 +\frac{C_2}{f_\pi^2} (N^\dagger\sigma
N)^2 +\cdots \label{L}
\ee
where ${\bf \pi}$ is the triplet pion field. 
Counting rules can be devised in such a way that one can
do a systematic expansion 
in some momentum scale $Q$ being probed that is in some sense small compared
to the typical chiral scale $\Lambda$ which can be taken to be roughly
of the mass of the heavy particles that have been
integrated out. If this is effectuated to
all orders, then according to the axiom given by Weinberg \cite{weinnature},
we are in principle doing a full theory.

Let us see what this scheme means for nuclear interactions. In so doing 
I will uncover what is called ``chiral filter phenomenon" in nuclear
processes. A physical amplitude involving $E_N$ external nucleon lines
can be written as
\be
A\sim Q^\nu F(Q/\Lambda)
\ee
where $F$ is a slowly varying function of the dimensionless quantity
$Q/\Lambda$. Given that $Q$ is small compared with the chiral scale,
the idea is to calculate to the highest order possible
in $\nu$ and sum the terms to
the order calculated. If one can do this to all orders, as mentioned,
then one is doing
the full theory. With the chiral Lagrangian that we are concerned with,
the counting rule can be readily deduced by looking at the Feynman diagrams.
One finds
\be
\nu=4-2C+2L-(E_N/2+ E_{ext}) + \sum_i V_i \bar{\nu}_i\label{index}
\ee
where $C$ stands for the number of clusters, $E_N$ the number of
external incoming and outgoing nucleon lines, $E_{ext}$ the number of
external fields, $L$ the number of loops and $V_i$ the number of vertices
of type $i$ and
\be
\bar{\nu}_i= d_i +\frac{n_i}{2} +e_i -2\label{subindex}
\ee
with $d_i$ the number of derivatives, $n_i$ the number of nucleon lines
and $e_i$ the number of external fields entering the $i$th vertex.

In this paper we will be concerned with at most one slowly varying external
field, so we will have $E_{ext}=1$. Obviously
 $e_i=1$ will appear only once if at all.
For two-body exchange currents, we are concerned with an irreducible graph
with $E_N=4$ and $C=1$. So the important quantities in (\ref{index}) are
the number of loops $L$ and the $\barnu$. For a given $L$, therefore,
only $\barnu$ matters.
The particular structure of chiral symmetry requires that
\be
\bar{\nu}_i\geq 0.
\ee

We can now state the ``chiral filter phenomenon" which will figure prominently
in what follows. 
In the form it was first stated \cite{kdr}, the general argument
that follows from effective chiral Lagrangians was not used. What the statement
says is that while nuclear forces involve long-range and 
short-range interactions on the same footing and hence have to be taken into
account at the same time, the response to a slowly varying electroweak
field {\it screens} short-distance physics, thereby causing  
the effect of soft-pion mediated
process to show up prominently (unless accidentally suppressed by kinematics). 
One can see this to the lowest order in the chiral 
counting. Take the two-nucleon potential generated by one-pion exchange
and the one generated by a contact four-Fermi interaction in Eq.(\ref{L}).
The one-pion exchange involves two vertices each of which has 
index $\bar{\nu}_i=0$ since there is one derivative in the pion-nucleon
coupling and two nucleons attached to the vertex, $n_i=2$. 
Higher-order terms in derivative will bring in higher
power and be suppressed. 
So at the zero-loop level, it is this tree term that matters.
But the same is true with the four-Fermi interaction with zero loops
as there is no derivative at the vertex but four nucleons enter with $n_i=4$, 
so again $\barnu=0$. The one-pion-exchange potential is
the longest-range one in two-nucleon systems. Now four-Fermi interactions
represent the short-range part of the potential in which heavy mesons 
representing the degrees of freedom lying above the chiral scale $\Lambda$ have
been integrated out. Thus we conclude that as far as the chiral counting 
is concerned both the {\it longest-range} and 
{\it shortest-range} potentials
contribute on an equal footing and cannot be
separated.

Consider attaching an external field to the same two-body system.
Attaching an external field to the one-pion-exchange term does not modify
the index $\barnu$ since $e_i+d_i=1$, hence we still have $\barnu=0$ in the
leading order but the four-Fermi interaction term requires one derivative
in addition to $e_i=1$ leading to an index
$\barnu\geq 2$. Therefore in contrast to the nucleon-nucleon potential,
short-range interaction terms are naturally suppressed relative to the
one-pion-exchange term
when attached to an external field\cite{mr91}. This is true for a slowly
varying electroweak current in general.

This ``chiral filtering" is both a good news and a bad news. It is
a good news in that meson-exchange currents can be under control
with the dominance of Goldstone pions without interference from 
poorly-understood short-range degrees of freedom. It
is a bad news since the pion dominance means that unless accidentally
suppressed, pions will not allow us to learn short-distance physics
through exchange currents. To some who would like to see a ``smoking 
gun" of quark-gluon degrees of freedom in nuclear processes, this is a pity.

\begin{figure}[tbp]
\setlength{\epsfysize}{4in}
\centerline{\epsffile{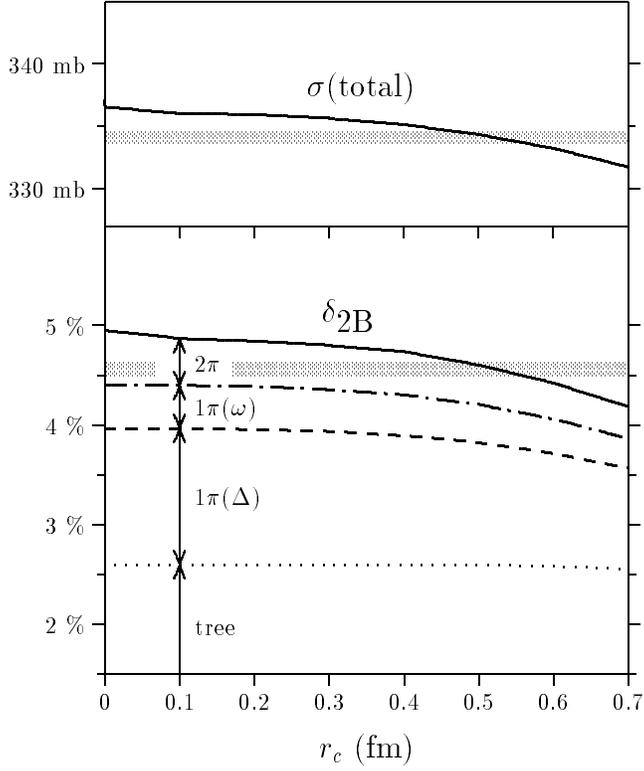}}
\caption{\protect \small
Total capture cross section $\sigma_{\rm cap}$ (top) and $\delta$'s (bottom)
vs. the hard-core radius $r_c$ that represents the uncertainty due to 
short-distance physics unaccounted for in low-order chiral perturbation theory.
The solid line represents the total
contributions and the experimental values are given by the shaded band
indicating the error bar.
The dotted line gives $\delta_{\rm tree}$, the dashed line
$\delta_{\rm tree} + \delta_{1\pi}^\Delta$, the dot-dashed line
$\delta_{\rm tree} + \delta_{1\pi}= \delta_{\rm tree} + \delta_{1\pi}^\Delta
+ \delta_{1\pi}^\omega$ and the solid line the total ratio, $\delta_{\rm 2B}$.}
\label{dataII}
\end{figure}

I show in Figure \ref{dataII} the results of the most recent calculation
to the next-to-next (NNO) order  in the chiral expansion \cite{pmr}.
I should explain in some detail what went into this calculation.
In calculating the cross section, imagine that we have the process occurring
between a proton and a neutron interacting to all orders through the chiral
Lagrangian of the form (\ref{L}). The initial state is the 
scattering state in $^1S_0$ and the final state is the bound (deuteron) 
$^3S_1$ state with a small D-state admixture. The electromagnetic
current -- which is predominantly magnetic dipole ($M1$) -- connects
the initial and final states. Both the bound state and the scattering state
with a large scattering length (with $a_s=-23.75$ fm) are not amenable to
a straightforward chiral perturbation expansion because of infrared divergences
and nobody knows how to calculate this in a systematic
chiral expansion. However this difficulty can be circumvented by 
realizing that what we are interested in is the meson-exchange current
contribution {\it relative} to the single-particle matrix element
for which we can take the most realistic wave functions available for the
initial and final states. Such wave functions are indeed in the market,
namely those computed from the Argonne $v_{18}$ potential \cite{v18}.
This procedure is not exactly a systematic chiral expansion since the
infrared-divergent reducible diagrams are summed to all orders (in
the form of solving the Schr\"odinger equation) while irreducible
graphs are computed to the NNL order only. However the NNL order 
essentially saturates the irreducible graphs within the uncertainty
associated with the short-distance part of the wave function, which is
of the order of less than 1\%, so the scheme is consistent as far as this
calculation is concerned.

To the extent that the wave functions are very accurate, the single-particle
matrix element will also be. One can gauge this by looking at the
prediction for the $^1S_0$ scattering length and the static properties of
the deuteron, all of which are remarkably accurately given
by the Argonne $v_{18}$ potential. 
It is therefore convenient to look only at the exchange-current
corrections {\it relative} to the single-particle matrix element. The dominant
one-pion exchange matrix element is ${\O} (Q^2)$ relative to the single-particle
matrix element. One-loop radiative corrections are further suppressed
by the same order. The ratios of the matrix elements relative to the
single-particle matrix elements are denoted $\delta$ in Figure \ref{dataII}.
The four-Fermi interactions found to be suppressed by the ``chiral filtering"
enter at the range of uncertainty associated with the short-range correlation
in the wave functions given by the hard-core radius $r_c$ in Figure 
\ref{dataII}. In fact, the suppression of the zero-range operators due
to the correlation function represents in an indirect way this part of 
physics\footnote{Work is in progress\cite{PKMR} to eliminate the ad hoc use of the hard-core
cutoff in the wave function by a systematic chiral expansion in the Hamiltonian
and in the wave function following a method recently discussed in a lecture by Lepage\cite{lepage}.}.
The $1\pi(\omega)$ and $1\pi(\Delta)$ correspond to one-pion exchange
contributions with vertex corrections due to 
the counter terms resonance-saturated with an $\omega$ and
a $\Delta$, respectively. The ${2\pi}$ corresponds to the 
two-pion exchange one-loop correction. The total 2-body correction is denoted
$\delta_{2B}$.

Now calculated
with the single-particle matrix element alone, the cross section comes out
to be
\be
\sigma_{imp}=305.6\ \ {\rm mb}
\ee
which differs by about 9\% from the experimental value 
\be
\sigma_{exp}=334.2\pm 0.5\ \ {\rm mb}.
\ee
The chiral Lagrangian treatment, taking into account the short-range 
uncertainty given above, gives an accurate result, accounting fully
for the missing 9\%,
\be
\sigma_{theory}=334\pm 3\ \ {\rm mb}
\ee
where the quoted error represents theoretical uncertainty associated with
short-distance physics.

Perhaps much less solid theoretically but more spectacular is the 
electrodisintegration
of the deuteron, a sort of inverse process to the np capture,
\be
e + d \rightarrow e+n+p.
\ee
If one applies the same formalism as in the np capture, it is found that
the meson-exchange current effect, while small in the np capture, becomes
big because of a substantial cancellation at finite momentum transfer
between the S-state and D-state components
of the deuteron wave functions. The chiral expansion to the next-to-next order
described above turns out to work well up to a large momentum transfer
of order $\sim 1$ GeV \cite{frois}. 
While it has not been checked in detail that other corrections
remain negligible here, the important 
presence of the mesonic current is clearly 
exhibited in this process. It remains to prove that higher order 
chiral corrections are indeed suppressed at large momentum transfers involved.

Another place where the chiral filtering is visibly operative
is in axial charge transitions in nuclei, i.e,
\be
J^+\leftrightarrow J^- \ \ \ \Delta T=1.
\ee
If $J=0$, this process is analogous to a pion decaying into the vacuum 
carrying an interesting information on the ``vacuum property" of
nuclear ground state. Warburton \cite{warburton}
studied extensively  this class of
transitions in light, medium and heavy nuclei, and obtained an important result that
as one goes to heavier nuclei the effect of the pion exchange
in the axial charge matrix elements becomes stronger. He defined
a quantity called $\epsilon_{MEC}$
\be
\mec= \frac{M_{exp}}{M_{sp}}
\ee
where $M_{exp}$ is the experimentally measured matrix element of the
axial charge operator and $M_{sp}$ is the theoretical matrix element
of the single-particle axial charge operator calculated with the
best possible nuclear wave functions available in the literature. 
Since this quantity involves both
experimental and theoretical quantities, it is not quite 
what one would call experimental value. There is an inherent uncertainty associated with
the single-particle matrix element.
What is however quite significant in the study of
Warburton is that unlike in the case of electromagnetic exchange currents
the effect of the chiral-filtered pions can be enormous. Indeed in light 
nuclei, $\mec$ is around 1.5, that is, the exchange correction is 50 \%
in the matrix element. This is a huge correction.
What is more 
significant is that in heavy nuclei, the effect is even more dramatic.
In lead region, Warburton found
\be
{\mec}=1.8\sim 2.0.\label{warburton}
\ee
The range is the uncertainty involved in the theoretical single-particle
matrix element alluded to above.

There is a simple way of getting the enhancement (\ref{warburton}). This
can be done by
combining the chiral filter mechanism together with
what is known as ``BR scaling"
in dense medium which I shall now explain. The idea involves once more
the general philosophy of effective Lagrangians but extrapolated further
into the regime where matter is dense and where direct measurements are
not readily available.

If physics does not change drastically from light to heavy nuclei, one
may start with a Lagrangian like (\ref{L}) and then compute systematically
the effect of the medium by suitably accounting for additional scales
brought in by matter density. There are efforts to do this sort of
calculations. Here I will consider approaching from the other extreme (say,
a ``top-down" approach)
where possible nonperturbative effects associated with the medium are taken
into account {\it ab initio}
in a manner consistent with the notion of chiral effective theories.
This has an advantage in that physics under extreme conditions such as
the state of dense matter encountered in compact neutron stars and relativistic
heavy-ion collisions can be treated on the same footing. Viewed in this way,
calculating the enhancement (\ref{warburton}) will be a low-order calculation
whereas starting from (\ref{L}) would require ``high-order" calculations.

In writing down Eq. (\ref{L}), I emphasized that it has particles whose 
parameters are defined in the absence of background matter. Now consider
a particle, a fermion or a boson, propagating in a medium consisting of
matter in a bound state like in the interior of a very heavy nucleus.
For this, I can start with a Skyrme type Lagrangian containing only meson
fields. Imagine having a realistic Lagrangian of such type containing not
only pions as in the original Skyrme Lagrangian but also vector mesons
and other heavy mesons. A nucleon with this Lagrangian comes out as a
soliton, ``skyrmion," with baryon number $B=1$ which is just the winding
number of the soliton. The same Lagrangian in principle
can describe the deuteron, triton
... and $B=\infty$ nucleus, all arising from the same Lagrangian. At present
we do not know how to write such a Lagrangian and hence we do not know
how to compute, for instance, the binding energy, 
the equilibrium density of nuclear
matter and nuclear matrix elements of currents. What is known is that
the deuteron and nuclear force can be reasonably understood even 
from a drastically simplified skyrmion Lagrangian\cite{CND}.

Given a realistic chiral Lagrangian of the skyrmion type, the question one can
ask is: How does a hadron propagate in a medium defined by a density
$\rho$, say ? The most obvious thing to do is then to write an effective
Lagrangian that has all the right symmetries of the original theory, QCD,
but suitable in the background defined in the presence of a medium.
In QCD, the quantity that reflects the background or the ``vacuum"
is the quark condensate
and since the background is changed, we expect that the condensate would be
suitably changed. Let me denote the modified condensate by putting an asterisk
\be
\la \bar{q}q\ra^\star\neq \la\bar{q}q\ra_0, \ \ \ \rho\neq 0.
\ee
Since the condensate is modified, all the associated quantities such
as light-quark hadron masses, the pion decay constant $f_\pi$ etc. will
be suitably modified. I will denote them with an asterisk on top. By following
the strategy of preserving the same symmetry present in matter-free
space {\it except that asterisked parameters enter}, it is possible to
establish the scaling\cite{BRscaling}
\be
m_V^\star/m_V\approx M_N^\star/m_N\approx m_\sigma^\star/m_\sigma\approx
f_\pi^\star/f_\pi\approx (\la \bar{q}q\ra^\star/ \la\bar{q}q\ra_0)^n\label{BR}
\ee
where the subscripts $V$, $\sigma$, and $N$ stand, respectively, for 
(light-quark) vector
meson, scalar meson and nucleon fields and the index $n$ is some power that
depends on specific models (for the simplest 
Skyrme model, $n=1$, for the NJL model,
$n=1/2$ etc.). An effective Lagrangian of the type (\ref{L}) with its
parameters given by (\ref{BR}) can then describe, at tree order, fluctuations
around the state defined by density $\rho$.

At present there is no systematic derivation of such an effective Lagrangian 
from first-principle arguments. As such, the scaling (\ref{BR}) is not a
relation that can be used in {\it any}
Lagrangian field theory dealing with nuclear matter.
It should be considered as a particular parameterization with a given Lagrangian
of the type I have been considering. Thus the quantities with such scaling can
have meaning only as {\it parameters} of a specific theory and it would be too hasty to identify them as ``physical'' masses and constants. The only 
quantity that is physically meaningful is the measurable one. 

One way to ``derive'' the scaled Lagrangian is to look for a non-topological
soliton of the effective action arising from a high order chiral perturbation
theory. As suggested by Lynn \cite{lynn}, it could be a ``chiral liquid'' that defines
the Fermi sea with a given Fermi momentum $k_F$. One can
identify this as a ``chiral-scale'' decimation in the renormalization
group approach mentioned below, with the cutoff set at the chiral scale
$\Lambda \sim \Lambda_\chi \sim 1$ GeV. Once such a ``chiral liquid'' state is
obtained, then the scaling parameters will be defined in fluctuations around
the chiral liquid state in what I would call ``Fermi-liquid scale decimation.''
I will return to this matter. 

Let us now go back to Warburton's $\mec$ in heavy nuclei for which we will
take $\rho\approx\rho_0$. Suitably coupling the axial current to
a BR effective Lagrangian, one can calculate and find \cite{pmraxial}
\be
\mec={\Phi}^{-1} (1+R)\label{mecth}
\ee
where
\be
\Phi:=f_\pi^\star/f_\pi\approx m_V^\star/m_V \cdots
\ee
and $R$ is the ratio of the matrix elements of the meson-exchange 
axial charge operator over the single-particle axial charge operator.
The meson-exchange operators are given in chiral perturbation theory
to the next-to-next-to leading order in the chiral expansion as in the
electromagnetic case, again dominated by the pions due to the chiral filter
as explained in \cite{kdr}. The ratio $R$ does not depend much on how one
calculates the matrix elements, that is, nuclear model-independent,
and depends only slightly on density. For heavy nuclei, it comes out
to be $R\approx 0.5\sim 0.6$. The quantity we need to compute $\mec$
is $\Phi$, the only quantity that knows that nuclear matter ``vacuum"
is different from the matter-free vacuum. There are two ways known to
get this quantity -- and this is not given by the strategy of effective chiral
Lagrangian field theory. One is to use the Gell-Mann-Oakes-Renner formula
for the pion embedded in nuclear medium, the other is to do a QCD sum-rule
calculation for the vector-meson mass in medium. While both quantities in medium are
not without ambiguity, they nonetheless give
the same answer. The result by the latter method
at $\rho=\rho_0$ is \cite{jin}
\be
m_V^\star/m_V=\Phi (\rho)=0.78\pm 0.08, \ \ \rho=\rho_0.\label{Phi}
\ee
With this value, (\ref{mecth}) gives
\be
\mec=1.9\sim 2.0.
\ee
This agrees with Warburton's analyses. As emphasized before, one could
of course calculate corrections to the chiral-filtered pionic contribution without
invoking BR scaling but instead using a vacuum-defined chiral Lagrangian and
explicitly incorporating other degrees of freedom (such as an effective scalar
meson $\sigma$ and light-quark vector mesons) and get the required enhancement
\cite{riska}. The two methods must be equivalent to leading order
at nuclear matter density.

It is easy to generalize the formalism to $SU(3)$ flavors and study
fluctuations in the strangeness directions. For instance, one could
look at the production of kaons in dense medium in heavy-ion collisions.
Once the ground state is defined in terms of BR scaling chiral Lagrangians,
fluctuations are then automatic at tree level, combining both flavor
$SU(3)$ symmetry and chiral symmetry. How this can be done is discussed in
\cite{SBMR}. Some of the predictions made in this way have been tested
by experiments recently performed at GSI (e.g., FOPI and KaoS) and are fairly well
confirmed\cite{llb97}. 
Extended smoothly beyond nuclear matter density, the theory can make
predictions on possible phase transition with condensation of kaons
at a density $\rho\sim 3\rho_0$ in compact-star matter like in nucleon
stars with a fascinating consequence on the formation of small
black holes and on 
the maximum mass of stable neutron stars etc. \cite{brownbethe}.

Up to this point, I have not discussed how the ground state, namely 
nuclear matter, comes out in this description. I shall now ``map" the
chiral Lagrangian with BR scaling treated in mean field to Landau
Fermi-liquid theory of nuclear matter developed by Migdal \cite{migdal}.
The idea is based on two observations. The first is that relativistic
mean-field theory for nuclear matter is known to be interpretable as
equivalent to Landau Fermi-liquid theory. For instance, Walecka's mean
field theory has been shown to be one such theory \cite{matsui}. The second 
is that Landau Fermi-liquid theory is a renormalization-group fixed point 
theory \cite{shankar}. This observation allows one to formulate a many-body 
problem from the point of view of effective field theories 
which is clearly what is needed
to go further into the unknown regime of high density.
 In a recent paper, Brown and I \cite{br96} argued that
the BR scaled chiral Lagrangian in a simplified form, when treated at
the mean field level, is equivalent to a Walecka-type mean field theory.
It is therefore quite logical that the BR scaled chiral Lagrangian
mean-field theory is equivalent to Landau Fermi-liquid fixed point theory.
The relevant arguments linking various elements of the theory are found
in \cite{SBMR}.

The crucial link is found at the stage of the second -- ``Fermi-liquid'' --
decimation that integrates out excitations of the scale $\tilde{\Lambda}$ around the Fermi surface defined by the Fermi momentum $\kF$ and then does the rescaling. 
The main ingredient is the renormalization group-flow result that
there are two fixed-point quantities in the theory\cite{shankar}. 
One of them is the effective
mass of the nucleon $m_N^\star$ and the other is the Landau interaction
${\cal F}$. Both are defined at the Fermi surface.
By Galilean invariance, the effective mass is related to the 
Landau parameter $F_1$ as
\be
\mnstar/\mn=1+\frac 13 \F1=\left(1-\frac 13 \tilde{F}_1\right)^{-1}
\label{landaumass}
\ee
where $\tilde{F}_1:=(\mn/\mnstar)\F1$. 
Now using that the Walecka model is equivalent
to Landau Fermi liquid, we deduce that $\tilde{F}_1$ 
gets a contribution from the
$\omega$ channel, say, $\ft1omega$. Due to chiral symmetry, there is also the Goldstone pion 
contribution through a Fock term to $\tilde{F}_1$ 
which can be explicitly calculated.
Thus
\be
\tilde{F}_1=\ft1omega +\tilde{F}_1^\pi
\ee
with
\be
\ft1omega=3(1-\mn/\Mnstar)=3(1-\Phi^{-1})\label{ft1omega}
\ee
and
\be
\tilde{F}_1^\pi
=-\frac{9f_{\pi NN}^2 \mn}{8\pi^2 \kF}\left[\frac{\mpi^2+ 2\kF^2}{2\kF^2}
\ln\frac{\mpi^2+4k_F^2}{\mpi^2} -2\right]\label{ft1pi}
\ee
where $\f_{\pi NN}\approx 1$ is the nonrelativistic $\pi N$ coupling constant.
Note that (\ref{ft1pi}) is precisely
determined once the Fermi momentum is given, say, $\approx -0.153$
at normal matter density. The important point here is that the effective mass
gets contributions from the (BR) scaling parameter (\ref{ft1omega}) {\it and}
the pion. The pion comes in as a perturbative correction to the nonperturbative ``vacuum''
contribution given by $\Phi$. The effective (Landau) mass 
(\ref{landaumass}) is therefore
\be
\mnstar/\mn=\left(\Phi^{-1}-\frac 13 \tilde{F}_1^\pi\right)^{-1}\label{Lformula}
\ee
which at $\rho=\rho_0$ predicts
\be
\mnstar/\mn (\rho_0)\approx 0.69.\label{effmass}
\ee
This is a genuine prediction which is supported by the orbital gyromagnetic
ratio in heavy nuclei, discussed below. 
It is also consistent with the QCD sum rule 
calculation of the nucleon mass in medium\cite{mnsr},
\be
(\mnstar/\mn)_{QCD} (\rho_0)=0.69^{+0.14}_{-0.07}.\label{mnqcd}
\ee

Now having the relation between the fixed point $\mnstar$ and $\Phi$ (plus
the calculable pionic term), we can derive
various interesting and highly nontrivial relations applicable to long-wavelength 
processes\cite{FR95}.
For instance, the EM convection current for a nucleon on the Fermi surface
which can be written down on the basis of $U(1)$ gauge invariance can
be derived from our chiral Lagrangian:
\be
{\bf J}=g_l \frac{\bf p}{\mn}
\ee
where $g_l$ is
the orbital gyromagnetic ratio given by
\be
g_l=\frac{1+\tau_3}{2}+\delta g_l
\ee
with
\be
\delta g_l=\frac 49 \left[\Phi^{-1} -1 -\frac 12 \tilde{F}_1^\pi\right]\tau_3.
\label{deltagl}
\ee
I should stress that this relation is highly non-trivial for several
reasons. First of all, the isoscalar current is given by ${\bf J}^{(0)}
={\bf p}/2\mn$, so the scaling mass $\Mnstar$ does not figure in the
current (this is an equivalent to ``Kohn theorem" in condensed matter physics)
and secondly the many-body nature of the system is manifested only in
the isovector part through $\delta g_l$. Given the numerical value (\ref{Phi})
at nuclear matter density, we get 
\be
\delta g_l=0.23\tau_3.
\ee
This should be compared with $\delta g_l^{proton}=0.23\pm 0.03$ obtained
from a dipole sum rule in $^{209}$Bi\cite{schumacher} and with $\delta
g_l^{proton}\approx 0.33$, $\delta g_l^{neutron}\approx -0.22$ obtained 
from an analysis of the magnetic moments in the $^{208}$Pb region\cite{yamazaki}. 

The deviation of the nucleon effective mass from the ``universal" scaling
factor $\Phi$, (\ref{Lformula}), arises from the presence of the Goldstone pions. In the skyrmion description,
the difference arises from the fact that aside from the known
current algebra term, an additional 
term -- Skyrme quartic term -- is needed for stabilizing the soliton 
that metamorphoses into the physical
nucleon. Expressed in terms of physical variables, the difference can be attributed to
the fact that the axial coupling constant $g_A$ can scale in nuclear medium\cite{BRscaling,FR95}.
It turns out that
\be
\frac{\mnstar}{\mn}=\left(\frac{g_A^\star}{g_A}\right)^{\frac 12} \Phi.\label{skyrmion}
\ee
Comparing with (\ref{Lformula}), we find that
\be
\frac{g_A^\star}{g_A}= (1+\frac 13 F_1^\pi)^2
=(1-\frac 13\Phi \tilde{F}_1^\pi)^{-2}.\label{gaskyrmion}
\ee
For nuclear matter ($\rho\approx \rho_0$), this predicts\footnote{Clearly
the formula (\ref{gaskyrmion}) cannot be valid beyond a certain density
$\gsim \rho_0$. It appears that $g_A^\star\approx 1$ is a fixed point.}
\be
{g_A^\star}(\rho_0)\approx 1.
\ee
This is quite close to what is found in nature\cite{gaexp}. The same result
was obtained many years ago in terms of the Landau-Migdal parameter
$g_0^\prime$ in the $\Delta N$ channel\cite{ROW} which has recently been 
interpreted as a counter term in higher order
chiral expansion\cite{PJM}. The relationship
between these different interpretations is not yet understood and remains to
be clarified.

This close agreement of the chain of predictions with experiments can be taken to confirm the
validity of the notion that the scaling (\ref{BR}) --
initially introduced as a vacuum change --
is associated with the Fermi liquid fixed point in many-body interactions. 
I will later use this observation to propose a dual description between QCD and hadronic
variables. More immediately this result should allow us to write down
an effective chiral Lagrangian with the scaling (\ref{BR}) 
that {\it in the mean-field approximation reproduces
exactly the above Fermi liquid structure}. The corresponding Lagrangian is\footnote{This
should be understood in the sense of the effective action in the mean field sense, 
with $\delta S^{eff}|_{\omega^\star,\phi^\star,...}=0$ for
$S^{eff}=\int \ d^4 x\ \L_{BR} (x)$. Fields not figuring in the mean field such as
pion field are not explicited here. However once the mean field is defined, fluctuations
into strange and non-strange directions can be described by
restoring pion, kaon,... fields
in a way consistent with chiral symmetry. Certain consistency conditions 
require that $g_v^\star/g_v$ scale like $\Phi$ whereas no such conditions
exist for the scalar constant $h$.}
\be
\L_{BR} &=& \bar{N}(i\gamma_{\mu}(\d^\mu+ig_v^\star\omega^\mu )-
M_N^\star+h\phi )N
\nonumber\\
& &-\frac 14 F_{\mu\nu}^2 +\frac 12 (\partial_\mu \phi)^2 
+\frac{{m^\star_\omega}^2}{2}\omega^2 
-\frac{{m^\star_s}^2}{2}\phi^2\label{toyBRPP}
\ee
where we have retained only the $\omega$ field and the effective scalar
field $\phi$ in the meson sector eliminating the pion field from the
chiral Lagrangian since we are to interpret it as an effective one to be
considered only in the mean field. I have written this Lagrangian in analogy to
Walecka's original linear $\sigma\omega$ model but it would be better to
consider it as a Lagrangian that obtains in a chirally invariant way from
one with 2-Fermi and 4-Fermi interactions using massive auxiliary fields
$\omega$ and $\phi$. Treated in the standard manner as for the Walecka model,
this effective Lagrangian describes nuclear matter fairly accurately. 
For instance
with the physical masses, $m_N=939$ MeV, $m_\omega=783$ MeV and $m_s=700$
MeV and the parameters, $h=6.62$, $g_v=15.8$ and assuming the
scaling
\be
\Phi(\rho)=(1+0.28\rho/\rho_0)^{-1}
\ee
normalized so that the known value (\ref{Phi}) is reproduced at $\rho=\rho_0$, 
we get the binding energy $B$, the equilibrium Fermi momentum $k_F$
and the compression modulus $K$:
\be
B=16.0\ \ \ {\mbox MeV}, \ \ \ k_F= 257\ \ \ {\mbox MeV}, \ \ \ K=296\ \ \
{\mbox MeV}.
\ee
The corresponding effective mass of the nucleon at the minimum is
\be
\mnstar=\Mnstar-h\la\phi\ra^{\star}=0.62 m_N
\ee
which should be compared with (\ref{effmass}) and (\ref{mnqcd}).
The nuclear matter property so obtained 
is quite close to that obtained from an effective chiral 
Lagrangian constructed based on {\it naturalness condition}\cite{furnstahl}.

The idea developed here allows one to explore what happens when matter is compressed
to a density greater than normal. This is a relevant issue for on-going
experiments in relativistic heavy-ion collisions and for understanding such compact stars as neutron stars. Suppose one would like to probe
the regime where $\rho > \rho_0$. Instead of approaching this regime
with a Lagrangian defined at $\rho=0$ as is done conventionally, I would
like to consider fluctuations 
around $\rho\approx \rho_0$ with the effective Lagrangian
{\it defined} at that point. The advantage in doing this is that even if fluctuating around
the $\rho=0$ vacuum were a strong-coupling process and hence required a high-order calculation, fluctuations around the ground state at $\rho=\rho_0$ could 
be weak-coupling allowing for a tree-order or at most a next-to-leading calculation.
Indeed the recent elegantly simple explanation of
the CERES dilepton data\cite{ceres} by Li, Ko and Brown (LKB) \cite{LKB}
is a nice example of such an application. Here one is probing hadronic matter
at a density $\sim 3\rho_0$ at some high temperature. In the
LKB approach, the dileptons measured in the experiments are interpreted as
arising from mesons in a heat bath with their masses scaled as
(\ref{BR}). The result is consistent with a quasiparticle picture 
for both nucleons and mesons in a heat bath. 

The argument developed to link effective chiral Lagrangians and Fermi-liquid theory 
is manifestly tailored for very low-energy 
excitations for which Landau quasiparticle picture is valid. For instance,
in describing nuclear matter ground state, the heavy meson fields whose parameters scale 
as (\ref{BR}) are way off-shell. In matter-free space their masses are comparable to the chiral
scale $\Lambda_\chi$ and hence one might naively think 
that processes involving excitations of such particles on-shell
could not be handled reliably by the argument based on chiral symmetry used here.
Now what is observed in the CERES experiment 
is highly excited modes, involving hundreds of MeV. In particular the
``$\rho$ meson'' which plays an important role in the description of \cite{LKB}
is near its mass shell {\it albeit at a scaled mass} and moving in the medium with certain momentum.
Thus it may be puzzling that the quasiparticle picture for the mesons works
so well. One would have expected that even within the given scheme, higher loop
graphs (e.g., widths) and explicit momentum dependences should enter importantly. This puzzle is 
further highlighted by the equally 
successful explanation of the same process by a description that
is based on standard many-body approach starting from a theory defined at zero density 
\cite{wambach} which relies on the mechanism that in medium, the width of the $\rho$ meson
increases\cite{KW} or the $\rho$ ``melts."  

One possible solution to this puzzle is that as alluded already,
the description based on BR-scaling chiral
Lagrangians and the one based on many-body hadronic interactions are ``dual" in the sense that
they represent the same physics\footnote{The duality I am referring to here
resembles in some sense the ``quark-hadron" duality much discussed in the
literature. For example, in the (1+1)-dimensional 't Hooft model,
one can describe the non-leptonic decay width of a heavy-light meson
(e.g., $B$ meson) in terms of either the sum of exclusive partial widths or
the tree-level partonic width. In the infinite mass limit of the meson,
the two descriptions are equivalent to each other, the difference arising
at the level of 1 over the mass\cite{duality}.}. 
What the CERES data are telling us is that this duality 
may be holding in the heat bath and that the two descriptions may be mapped to each other\cite{geb}. This may be understood in terms of a ``mended symmetry"\cite{mended}. As interpreted in \cite{klr}, the mended symmetry argument goes as
follows. While in matter-free space, chiral symmetry is non-linearly realized with
the massive scalar degree of freedom purged from the low-energy sector, as density
increases, a scalar, say, the $\sigma$, comes down to join the triplet of the pion
to ``mend" the
$O(4)$ symmetry of the chiral $SU(2)\times SU(2)$ and to become the fourth component of the
four vector. That is, in dense medium, the non-linear $\sigma$ model is 
``mended" to
the linear $\sigma$ model with the masses BR-scaling\cite{klr}.
How this can happen in nuclear dynamics with the broad scalar in 
matter-free space becoming a local field in dense medium is described in \cite{brPR}. Now the
vector mesons, as long as they are still heavy, can be introduced much like the nucleon
as matter fields with their
masses scaling as (\ref{BR}). As density increases further and approaches
the critical density for the chiral transition, then the vector mesons become light and the Georgi
vector symmetry\cite{georgi,brPR} would be ``mended." {\it This interpretation clearly puts more significance on the symmetry consideration
than on the complex dynamics (e.g., used in \cite{wambach}), in 
conformity with what has been established in QCD at
long wavelength, namely, that in low-energy regime, it is chiral symmetry of QCD that 
governs the
physics of hadrons}. I suspect that it is this aspect that is at the root of the duality
we see in the CERES data. It would be extremely interesting to see
whether this dual description continues to hold 
true when heavy mesons are probed in cold nuclear matter 
as in Jefferson Laboratory or denser (somewhat warm) matter as 
in HADES.

\subsubsection*{Acknowledgments}

Part of this note was written while I was visiting Departament de Fisica Te\`orica, 
Universitat de Val\`encia, Burjassot (Val\`encia), Spain as an IBERDROLA Visiting Professor.
I would like to thank Vicente Vento and other members of the theory department for 
the pleasurable stay.

\end{document}